\documentclass[12pt]{article}

\usepackage[english]{babel}
\usepackage[utf8]{inputenc}
\usepackage{mathtools} 
\usepackage{array} 

\usepackage{graphicx}
\usepackage[colorinlistoftodos]{todonotes}
\usepackage{booktabs}
\usepackage[margin=1in]{geometry}
\usepackage{indentfirst}
\usepackage{float}
\usepackage[compact]{titlesec} 
\usepackage{soul}
\usepackage{xcolor}
\usepackage{caption} 
\usepackage[left, running]{lineno} 

\DeclareUnicodeCharacter{2212}{-} 

\newcommand{\beginsupplement}{%
	\setcounter{section}{0}
	\renewcommand{\thesection}{S\arabic{section}}%
	\setcounter{equation}{0}
	\renewcommand{\theequation}{S\arabic{equation}}%
	\setcounter{table}{0}
	\renewcommand{\thetable}{S\arabic{table}}%
	\setcounter{figure}{0}
	\renewcommand{\thefigure}{S\arabic{figure}}%
}

\usepackage[numbers]{natbib} 

\setcitestyle{aysep={}} 
\usepackage{etoolbox}
\usepackage{filecontents}
\newbool{MyRefNumbers}
\booltrue{MyRefNumbers} 

\usepackage{url}
\usepackage{breakurl}

\usepackage{setspace}

\begin{document}
	\begin{center}
		\begin{Large}
			A well-timed switch from local to global agreements accelerates climate change mitigation \\
		\end{Large}
	\end{center}
		Vadim A.\ Karatayev$^1$*, V\'itor V.\ Vasconcelos$^{2,3,4,5,6}$, Anne-Sophie Lafuite$^1$, Simon A.\ Levin$^{3,5,7,8}$, Chris T.\ Bauch$^9$, Madhur Anand$^1$ \\  
\vspace{-4pt}

$^1$ School of Environmental Sciences, University of Guelph, Guelph, Canada.
$^2$ Princeton Institute for International and Regional Studies, Princeton University, Princeton, NJ, USA.
$^3$ Department of Ecology and Evolutionary Biology, Princeton University, Princeton, NJ, USA.
$^4$ Andlinger Center for Energy and the Environment, Princeton University, Princeton, NJ, USA.
$^5$ Princeton Environmental Institute, Princeton University, Princeton, NJ, USA.
$^6$ Informatics Institute, University of Amsterdam, Amsterdam, Netherlands.
$^7$ Resources for the Future, Washington, DC, USA.
$^8$ Beijer Institute of Ecological Economics, Stockholm, Sweden.
$^9$ Department of Applied Mathematics, University of Waterloo, Waterloo, Canada. 

* - corresponding author. E-mail vkaratay@uoguelph.ca

\par \textbf{Abstract:} Recent attempts at cooperating on climate change mitigation highlight the limited efficacy of large-scale agreements, when commitment to mitigation is costly and initially rare. Bottom-up approaches using region-specific mitigation agreements promise greater success, at the cost of slowing global adoption. Here, we show that a well-timed switch from regional to global negotiations dramatically accelerates climate mitigation compared to using only local, only global, or both agreement types simultaneously. This highlights the scale-specific roles of mitigation incentives: local incentives capitalize on regional differences (e.g., where recent disasters incentivize mitigation) by committing early-adopting regions, after which global agreements draw in late-adopting regions. We conclude that global agreements are key to overcoming the expenses of mitigation and economic rivalry among regions but should be attempted once regional agreements are common. Gradually up-scaling efforts could likewise accelerate mitigation at smaller scales, for instance when costly ecosystem restoration initially faces limited public and legislative support. \\


Existing attempts to realize international cooperation on halting climate change highlight that many ecosystem services are public goods that can incentivize beneficiaries to free-ride on others’ mitigation efforts \citep{OstromEA02}. The limited capacity of the atmosphere to absorb greenhouse gases before emissions severely affect climate globally is one public good that can be preserved though investments to reduce emissions (`mitigation' hereafter) \citep{DolsakO03,TavoniL14}. Adopting alternative investment opportunities with faster payoffs however can delay cooperation unless it is reinforced by social norms \citep{Levin14} or short-term climate forecasts become severe \citep{BuryEA19}. Players can also delay mitigation to wait for others to develop mitigation technology and infrastructure, or to wait for rival players to invest first and give up their competitive advantage \citep{FalknerEA10}. At the same time, achieving climate change mitigation globally and rapidly is critical to minimizing adaptation costs, and in preventing irreversible ecosystem degradation (e.g., species loss, \citep{CeballosEA20,TrisosEA2020}), degradation of human well-being, especially in developing countries that have lower capacity for climate adaptation \citep{AdgerEA03,MertzEA09}, and preempting the collapse of favorable climate regimes when human impacts exceed environmental tipping points \citep{LentonEA08,LentonEA19}.

Advancing mitigation commitments requires incentives such as trade agreements or laws \citep{FischerN08} (`agreements' hereafter) that are specific to the spatial scales of political bodies. The failure of global climate agreements to be effective has led to a focus on building blocks approaches to global cooperation, in which an agreement grows from local setups \citep{OstromEA02,StewartEA13,Cole15}. Game-theoretic work predicts that such localized mitigation agreements might succeed where global cooperation fails by reducing the number of interacting players and the potential for free-riding \citep{Ostrom90,Ostrom10,SantosP11,VasconcelosEA13}. 
%
However, the greater potential for bottom-up approaches to achieve mitigation locally can trade off with a much longer time to achieving mitigation globally, as change happens more slowly in more distributed systems \citep{DietzEA03}. 
So, can societies enhance both the probability and pace of desired outcomes, such as reducing greenhouse gas emissions, by a properly-timed switch from local- to large-scale mitigation agreements?

To answer this question, we consider a system of $L$ regional groups, representing coalitions of countries with shared interests or geography (e.g., BRICS, AILAC, Western Climate Initiative), formal unions (e.g., EU), or countries with partly-independent states (e.g., United States). Given that many benefits of climate mitigation require global cooperation and currently are politically debated, we focus on the impact of the (high) costs of mitigation. In each region, players can free-ride by not acting on the provision of the public good, or they can contribute to providing the public good by mitigating emissions. Simultaneously, mitigators support agreements that, when sufficiently endorsed, punish free-riding through fines, sanctions, or other penalties. Such sanctioning agreements can be local, affecting players only within a specific region, global, supported by and affecting all players across regions, or both \citep{VasconcelosEA20}. We examine whether bottom-up regional (local) agreements, world-wide (global) agreements, or a dynamic combination of these two best promotes fast climate change mitigation.

\textbf{System dynamics.} As observed in other research \citep{VasconcelosEA13,DannenbergG19} and in reality \citep{WalkerEA09}, we find that premature global-scale efforts to achieve cooperation on climate invariably fail (Fig.\ 1a). Although exogenous events occasionally produce brief mitigation efforts within individual regions, at a global scale, mitigation never reaches the quorum required to institute sanctioning agreements (Fig.\ 1b.iii). In contrast, local sanctioning agreements can capitalize on stochastically induced, regional mitigator dominance to establish mitigation as the norm within individual regions. Eventually, this process occurs in every local group, translating to global climate mitigation (Fig.\ 1b.ii).

Once local agreements produce a sufficiently high global proportion of mitigators, however, a shift to global agreements more than doubles the rate of mitigation spread (Fig.\ 1b, 2a). This difference arises because enacting global sanctions once mitigation is common pushes non-mitigating regions to cooperate on reducing emissions. Compared with local agreements only, switching to global agreements doubles the amount of accumulated mitigation achieved by year 60 (Fig.\ 1a) and halves the time it takes for 80\% of countries to commit to mitigation (Fig.\ 2a). This benefit of switching to global agreements increases with mitigation cost, sanctioning cost, and the role of payoffs in strategy choice (see Supplementary Figure 1 for sensitivity analysis) because players quickly reconsider mitigation commitments that arise through chance events.
High mitigator costs therefore decrease the overall probability that chance events induce mitigation without the benefit of among-region sanctions that occurs through global agreements once many regions commit to mitigation. Notably, the threshold required for a switch to be effective does not need to exceed a 50\% majority of regions (Fig.\ 1a).

Bootstrapping cooperation through local agreements may face difficulties. Investments in a shared technology, infrastructure, or public support create impediments to early motivations. Typically, these investments are initially more costly and only as they are implemented they become progressively more accessible, as more nearby players become mitigators. This occurs through research and development of industries of trading partners and industrially related products \citep{HidalgoEA07}, and economies of scale (we refer to these effects collectively as economies of scale). When there are substantial gains for scaling, even though initial costs are high, we observe that such economies of scale further increase the benefits of switching to global agreements (19\%, Fig.\ 2b) as inter-group interactions pull along non-mitigating regions where mitigation costs remain high, making regional action even more important.

A second major obstacle in climate change cooperation arises at the global scale from game-of-chicken dynamics among economically competing regions, for which investment in mitigation can reduce the competitiveness of regional economies through opportunity costs (Falkner et al. 2010). For instance, China maintains that industrialized countries should invest in mitigation first so that other countries can catch up economically, while industrialized countries such as the United States wish to retain a competitive advantage. Accounting for such inter-region rivalry, we find that group rivalry disproportionally slows down mitigation through locally based agreements. As local agreements provide no incentive for pairs of intensely competing regions to commit to mitigation, a switch to global institutions is more critical in pulling along holdout regions (Fig.\ 2c). Unlike economies of scale operating within regions, the positive feedback by which group rivalry holds back mitigation arises at the global scale. This distinction is highlighted by the fact that, with local agreements only, some regions experiencing strong competition never adopt mitigation. This highlights that global agreements can play a key role in promoting not only the pace but also chance of global climate cooperation.

\textbf{Importance of switching agreement scales.} Given that both local and global agreements are critical to promoting mitigation, does a concerted strategy attempting agreements at both scales simultaneously improve outcomes or does it risk diluting limited resources? Attempts to form sanctioning agreements can involve constitutive costs – resources spent even if agreements fail to establish, such as lobbying for a carbon tax bill that ultimately fails. Likewise when enacting timely and effective laws is not possible, sanctioning might initially depend on a few audacious players that refuse cooperation with non-mitigators at substantial cost to themselves, even if sanctions ultimately fail to promote mitigation. For instance, trade sanctions by Western European nations against all nations lacking aggressive climate mitigation commitments might disproportionally weaken Western European economies when global mitigation commitment is rare.

We find that under high constitutive sanctioning costs mitigators must first focus all efforts locally, for instance refusing trade only with non-mitigators who are in their group (Fig.\ 3). Thereafter, we find that the switch from local to global mitigation incentives is best done earlier (i.e., lower switching threshold $T$) when agreements are easier to establish (i.e., require less support or smaller quorums to effectively sanction defectors) or global agreements have greater efficacy (Supplementary Figure 1), but premature shifts to global agreements can delay mitigation at low levels for extended periods (Fig.\ 1b.iii).
For low constitutive sanctioning costs, efforts to establish local and global sanctioning agreements simultaneously do little to affect outcomes (Fig.\ 3) because global agreements initially fail and local agreements eventually become redundant. However, a more graduated shift where local and global agreement attempts overlap might be more beneficial when early global dialogue establishes a framework for later negotiations, when sanctions are more effective among neighbors than among regions, or when uncertainties exist in players' commitments to mitigation or in the quorum needed to establish a global agreement.

\textbf{Agreements have scale-specific roles.} We find that a timely shift from local to global agreements consistently accelerates and establishes global climate mitigation over agreements focused on a single scale. 
Building on analyses of cooperation dynamics near steady-state \citep{VasconcelosEA13}, our focus on transient dynamics highlights that global agreements play a key role in pulling along late-adopting regions once mitigation is sufficiently common. This role increases as economies of scale, among-group rivalry, or high costs (Supplementary Figure 1) make mitigation less likely to arise through chance events such as popular movements or natural disasters.
Global agreements investigated here can also complement a regime complex approach to mitigation, wherein groups can form over different mitigation targets and address different issues that collectively tackle climate change \citep{KeohaneV11,HannamEA17}.
Overlap in group membership can accelerate mitigation globally by emphasizing the marginal gains to cooperation when information about the outcomes of action is missing \citep{VasconcelosEA20}.
Therefore the current reality that players commit to multiple mitigation targets and political coalitions, some of which enjoy a high membership, could further incentivize mitigation and reduce the extent of sanctioning that falls upon local and global agreements.

Our multi-scale model also underscores the importance of local agreements in establishing mitigation during the early phases of conservation efforts. First, local agreements can capitalize on increases in mitigation that arise from local variation but have little impact on mitigation at the global scale, as seen in lower benefits of switching as stochasticity becomes more prevalent in strategy choice (see SM for description of the mitigation behavior for varying group sizes, rate of exogenous events and impact of payoff). This aspect can be especially critical as climate disasters create strong but region-specific momenta for climate mitigation \citep{Weber10,BeckageEA18}. For instance, whilst recent wildfires in Australia reinforce climate change concerns globally, they could create an especially strong case for changing the domestic policies in Australia towards mitigation. Second, local agreements ‘lock-in’ mitigation as the norm in early-adopting regions during early adoption phases when global agreements fail. For this reason, local agreements help mitigation establish and spread when rare whereas global agreements fail \citep{VasconcelosEA13}.

\textbf{Implications for ecosystem restoration.} While we focus on mitigating climate change on the international stage, the benefit of leveraging the scale-specific roles of agreements found here can also accelerate environmental mitigation at smaller scales. In addition to climate negotiations, mitigating the impacts of climate disasters and reducing greenhouse gas emissions will increasingly require restoring ecosystems over large spatial scales \citep{AronsonA13}. The urgency of ecosystem-level mitigation rises further as concentrated human impacts can be slow to change \citep{TekwaEA19} and ongoing ecological feedbacks are accelerating ecosystem degradation \citep{FloresEA17}. As with climate change mitigation, ecosystem restoration can be costly when local economies must limit resource extraction or reorient towards recreation. Initial public support can therefore be limited to a few local communities where traditional values, public outreach, or successful restoration projects establish conservation as a self-sustaining social norm \citep{NyborgEA16,Reyes-GarciaEA19}.

Building on this, our results suggest that environmental agencies first focus limited resources locally rather than in many places at once, prioritizing communities where restoration has lower costs or greater public support. Establishing ‘buy-in’ among local communities can in turn accelerate and sustain ecosystem restoration \citep{MadzudzoEA06,MorrisonEA09,CetasY17}. Ecological feedback loops can further increase importance of focusing efforts by creating threshold effects, reversing meager restoration efforts while increasing resilience of restored locations to disturbance \citep{SudingEA04}. After a string of local successes garners greater public support, a focus shift towards enacting state-level legislation and concerted restoration efforts in all remaining locations could ultimately boost the resources, public support, and efficiency of restoration projects \citep{NeesonEA15,AronsonA13}. While most successful conservation movements begin with activism and culminate in legislation, our results underscore the importance of focusing efforts on one scale at a time and carefully considering when to shift from local to regional restoration and public outreach. \\

Continually worsening climate will eventually force countries to reduce greenhouse gas emissions, with current debates centering on the urgency and best course of action to achieve global mitigation. We have shown one strategy that can greatly accelerate cooperation to mitigate climate change in the near future by leveraging local and global efforts.

\vspace{5pt}
\section*{Methods}
We consider $L$ regions. Each region, $j$, contains $N$ players, of which $M_j$ pay a substantial cost $c$ to reduce emissions while all others do not mitigate, for a mitigator proportion $m_j=M_jN^{−1}$. When mitigators reach a majority, consensus establishes agreements that sanction non-mitigators through fines or economic sanctions by an amount $p_{NM}$. Mitigators also incur a substantial cost of sanctioning ($p_M < p_{NM}$), for instance when sanctions prohibit lucrative trade opportunities, a fraction $f$ of which is incurred even if non-mitigators are not sanctioned when sanctioning coalitions, institutions, or bills fail to establish.

We consider a combination of local and global efforts where mitigators initially form only local agreements until mitigation proportion $m_G$ exceeds a threshold $T_L$, and then shift their efforts towards global agreements once global mitigation proportion $m_G$ exceeds a threshold $T_G \leq T_L$ (when these are the same we write $T$). We consider four types of agreements: i) coordinated flip from local to global agreements, $0 < T < 1$; ii) purely local agreements, $T=1$; iii) purely global agreements, $T=0$; and iv) temporary coexistence of local and global agreements, $0 < T_G < T_L < 1$. Global agreements might additionally differ in sanctioning efficacy by a factor $E$. At each time step, one randomly chosen player decides to reconsider their climate strategy with probability $\kappa$, the social learning rate. Players tend to choose strategies with greater payoffs to a degree $\beta$, which reflects the importance of financial considerations in climate policy. With probability $\mu$, however, chance events supersede any payoff differences and players choose strategies at random.

Throughout, we use a decreasing sigmoid function $\Theta(x) = (1+\exp(h x))^{-1}$ to model smooth transitions from local to global agreements ($h=20$, $0 \leq x \leq 1$). 
We also allow the point at which local agreements are phased out to occur well after global agreement attempts begin, with local and global agreement attempts (and, potentially, sanctioning of non-mitigators) happening one at a time for $b=0$ and overlapping throughout the mitigation process for $T=0$ and $b>1$.
Thus, the amount of punishment attempted by mitigators is $\delta(m_j) = \Theta(m_G-T-b) + \Theta(T-m_G)$ and reflects a switch from local to global agreement attempts as $m_G$ exceeds $T$.
The amount of punishment realized by successfully established agreements is then
\begin{linenomath}
\begin{align}
\Delta(m_j) &= \Theta(m_G-T-b) \Theta(p_T-m_j) + \Theta(T-m_G) \Theta(p_T-m_G) E.
\end{align}
\end{linenomath}
The total payoffs for strategy $s$ can depend on global ($m_G= (LN)^{-1}\sum_j M_j$) and local ($m_j=M_j N^{-1}$) mitigation frequency:
\begin{linenomath}
\begin{align}
\Pi_{NM}(m_j) &= - p_{NM} \Delta(m_j) \\
\Pi_{M}(m_j) &= - c - p_M (f\delta(m_j) + (1-f) \Delta(m_j)).
\end{align}
\end{linenomath}
At each model iteration, a randomly chosen player in group $j$ with strategy $s$ switches to strategy $k$ (with $n_{k,j}$ denoting the number of players with strategy $k$) with probability
\begin{linenomath}
\begin{align}
\Pr_{s\rightarrow k,j} =& \kappa \biggl( \dfrac{(1-\mu)n_{k,j}N^{-1}}{1+\exp(\beta(\Pi_{s}(m_j)-\Pi_{k}(m_j)))} + \mu \biggr).
\end{align}
\end{linenomath}

To model economies of scale, we set initial mitigation costs $c_0$ that decline by a fraction $r_L$ past $m_j=r_T$ using the updated mitigation cost $c(m_j) = c_0 (1-r_L \Theta(r_T - m_j))$, where $c_0=c (r_T + (1-r_T) (1-r_L))^{-1}$ ensures that $\int_{m_j=0}^{1}c(m_j) = c$ in our base model.

To account for the effects of among-region economic competition, we model reduced payoffs for regions where players commit to climate mitigation \textit{in leu} of alternative investments to bolster their economies.
Among-group interactions in the matrix $\textbf{A}$ are $0$ for $a_{i=j}$ while $a_{i\neq j}$ terms are drawn from a multivariate uniform distribution with rivalry reciprocity $R=cor(a_{i,j},a_{j,i})>0$.
Given group-level payoffs as $P(\vec{m})=(\vec{m} \Pi_{M}(\vec{m}) + (1-\vec{m}) \Pi_{NM}(\vec{m}))(\textbf{A}+\textbf{I}_L)$, we normalize the rows of $\textbf{A}$ by their sum to obtain $\textbf{A}^N $ and arrive at the fitness of group $j$ with $n$ mitigators $\lambda_{j,n} = P(\vec{m'}) \cdot (\textbf{A}^N-\textbf{I}_L)_j $, where $\vec{m'}_{i=j}=n$ and $\vec{m}_i$ otherwise.
We then incorporate the consequences of individual choice for group-level payoffs using the updated payoffs $\Pi_s^G(m_j) = \Pi_s(m_j) + \alpha G_s /2$, where $G_{NM}=\lambda_{j,m_j-N^{-1}} - \lambda_{j,m_j}$, $G_M=\lambda_{j,m_j+N^{-1}} - \lambda_{j,m_j}$, and $\alpha$ scales individual payoff of group competitiveness relative to punishment and mitigation costs.
With this formulation mitigation adoption slows when intense competition happens among pairs of non-mitigating regions.

\section*{Acknowledgments}
We thank Easton White and Mikaela Provost for feedback that improved the manuscript. This research was supported by the Natural Sciences and Engineering Research Council and the New Frontiers Research Fund (to M.A. and C.T.B.).

\section*{Author Contributions}
V.A.K. conceived the study, V.A.K., C.T.B., V.V.V., and M.A. designed and analyzed the model, and all authors participated in writing the manuscript.

\section*{Competing Interests}
The authors declare no competing interests.

\bibliographystyle{nature} 
\bibliography{savedrecs_hierarchy} 

\vspace{60pt}

\begin{figure}[H]
	\centerline{\includegraphics[scale=0.75,trim=2 .2 .1 2,clip]{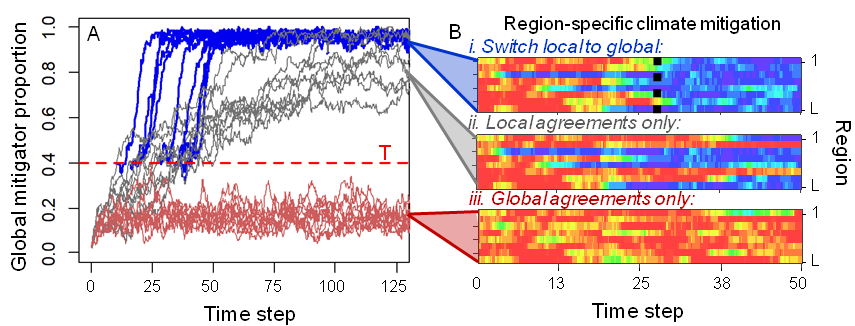}}
	\centering
	\label{fig:Hier0,4Fig1}
	\caption{Mitigation is best achieved through a timely switch from local to global agreements, on account of scale-specific mechanisms.
		(A) Time series of the proportion of system-wide climate mitigation over 10 replicate model runs with either global agreements only (red), local agreements only (gray), or a switch from local to global agreements (blue lines) once mitigation is sufficiently common (threshold $T=0.5$).
		(B) Region-specific proportions of mitigators from 0 (red) to 100\% (blue) within a single model run under three approaches. Unlike global agreements (B.iii), local agreements are critical for establishing mitigation as the norm in early-adopting regions (e.g., region 3 in B.ii); global agreements spread mitigation across regions once it is sufficiently common (B.i, dashed line denoting switch to global at $T$).
		Simulations show 8 regions with 18 players in each and no among-region rivalry or economies of scale; all other parameters values listed in Supplementary Figure 1.
	}
\end{figure}

\begin{figure}[H]
	\centerline{\includegraphics[scale=.22,trim=.2 .2 .1 2,clip]{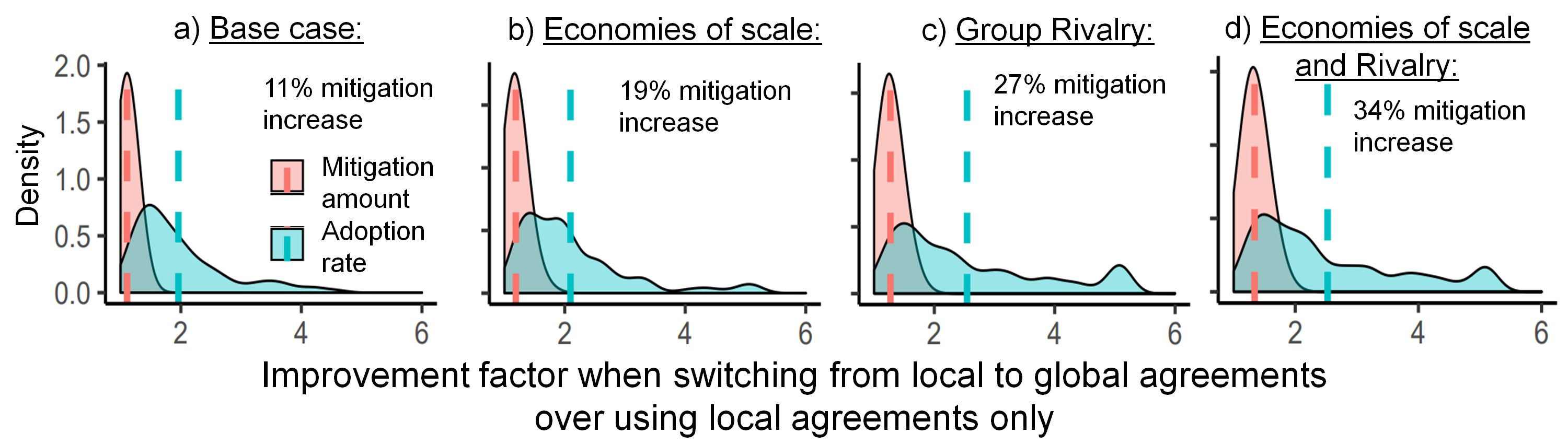}}
	\centering
	\label{fig:Hier0,35Fig2}
	\caption{Benefit of switching from local to global agreements increases when economies of scale and group rivalry affect mitigation. Values denote the factor by which switching outperforms local-only approaches (i.e., 2=200\% increase) in mitigation adoption rate (teal) and the total amount of mitigation achieved by time step 625 (red). Distributions give relative improvements in rates and total mitigation found within each of 100 replicate model runs, and vertical dashed lines denote mean improvement in each metric. Note that switching outperformed local-only strategies in all cases (i.e., all values $>1$). Economies of scale excluded in (a, c), group rivalry excluded in (a, b), and switch from local to global strategies happens at $T=0.5$; all other parameters values listed in Supplementary Figure 1.}
\end{figure}

\begin{figure}[H]
	\centerline{\includegraphics[scale=.75,trim=.2 .2 .1 2,clip]{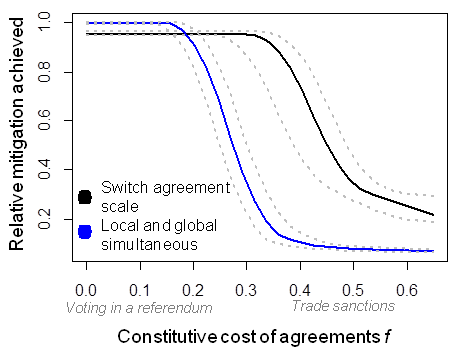}}
	\centering
	\label{fig:Hier0,35Fig4}
	\caption{Switching agreement scale (black, $b=0$) outperforms simultaneous local and global agreement attempts to promote mitigation (blue, $b=1.5$) above low constitutive agreement costs $f$. $f$ denotes the fraction of mitigator costs $p_M$ incurred at all times, regardless of whether sanctioning is in effect. Solid lines denote means and dotted lines denote the 25th and 75th quantiles in mitigation achieved by time step 625 across 30 replicate trials.}
\end{figure}

\newpage
\beginsupplement

\begin{large}
	\begin{center}
		{\bf Supplementary Information} \\
	\end{center}
\end{large}

\begin{figure}[H]
	\centerline{\includegraphics[scale=.62,trim=.2 .2 .1 2,clip]{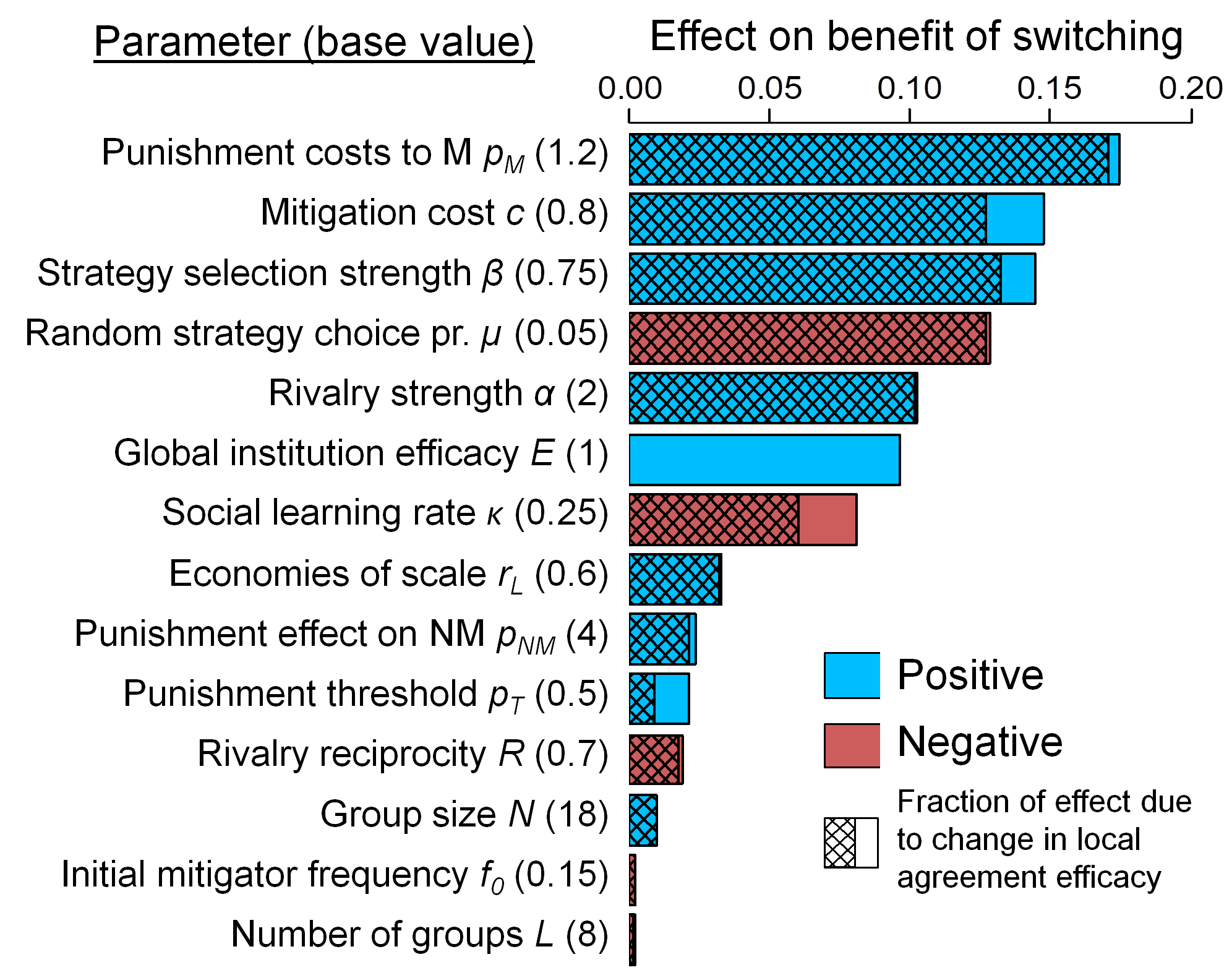}}
	\centering
	\label{fig:Hier0,35Fig3}
	\caption*{\textbf{Supplementary Figure 1.} Drivers of the benefits of switching from local to global agreements over local agreements only.
		Blue (red) bars denote greater (lower) benefits of switching with an increase in each parameter from 80\% to 120\% of its base value.
		Hashes on each bar denote the proportion of each change in switching benefit caused by a change in mitigation adoption rate using local agreements only, and non-hashed areas denote the proportion of each change in switching benefit caused by a change in mitigation adoption rate after global agreements begin.
		We quantify switching benefits as the proportional increase in total mitigation over the first 625 time steps over 120 replicates.}
\end{figure}

\end{document}